\documentclass[aip,graphicx]{revtex4-1}
\usepackage{graphicx}% Include figure files
\usepackage{dcolumn}% Align table columns on decimal point
\usepackage{bm}% bold math

\draft % marks overfull lines with a black rule on the right

\begin{document}

%\preprint{}

\title{Characteristic noise features in light transmission across membrane protein undergoing photocycle} %Title of paper



\author{Anshuman. J. Das}
\author{Sabyasachi Mukhopadhyay}%
\author{K. S. Narayan}
 \email{narayan@jncasr.ac.in}
\affiliation{%
Jawaharlal Nehru Centre for Advanced Scientific Research, Jakkur, Bangalore-560064, India%\\This line break forced with \textbackslash\textbackslash
}%


\date{\today}% It is always \today, today,

\begin{abstract}
We demonstrate a technique based on noise measurements which can be utilized to study dynamical processes in protein assembly. Direct visualization of dynamics in membrane protein system such as bacteriorhodopsin (bR) upon photostimulation are quite challenging.  bR represents a model system  where the stimulus-triggered structural dynamics and biological functions are directly correlated. Our method utilizes a pump-probe near field microscopy method in the transmission mode and involves analyzing the transmittance fluctuations from a finite size of molecular assembly. Probability density distributions indicating the effects of finite size and statistical correlations appear as a characteristic frequency distribution in the noise spectra of bR whose origin can be traced to photocycle kinetics. Valuable insight into the molecular processes were obtained from the noise studies of bR and its mutant D96N as a function of external parameters such as temperature, humidity or presence of an additional pump source.\end{abstract}

\pacs{Valid PACS appear here}% PACS, the Physics and Astronomy
%Use showkeys class option if keyword
\maketitle

\textbf{To appear in the Journal of Chemical Physics, Vol. 134, Issue 6}
\section{INTRODUCTION}
%\label{}
Noise features from a functioning system can reveal valuable insight into processes occurring over several time scales\cite{chen_prl}. Stochastic fluctuations are observed in systems like gene expression, protein, nano-medicine and can be used to probe and characterize gene circuits and phase transitions\cite{raser_sci,simpson_nano}. Identifying sources of noise, controlling and processing noise-related data using sophisticated techniques based on frequency domain analysis provides insight into molecular information that control the system\cite{cox_pnas,rao_nat,grima_jcp}. Noise as a tool to study electronic events and processes at microscopic length scales is extensively utilized in condensed matter physics. There are not many situations encountered where noise studies of optical absorption processes at room temperature has proven to be useful. Noise in biological processes are dealt at macroscopic-systems level and their correlation to molecular events are rarely emphasized. It is expected that fluctuations in a small system to be pronounced and we utilize that aspect by choosing a model biomolecular organization with the appropriate system parameters. Typical fluctuations from finite sized and correlated systems yield probability density functions (PDF) that deviate from Gaussian or Lorentzian distributions\cite{montroll_pnas, hill_jpc}. Kinetics associated with biochemical reactions can be modeled based on modified Langevin equations (LEQ). Solution of LEQ expressed in terms of reaction rates for mRNA and protein systems has been used to obtain noise frequency ranges which in turn are related to the decay rates\cite{simpson_jtb}. 
It has also been shown that fluctuations in small systems are larger in magnitude and can get comparable to the mean values of the variables\cite{simpson_nano}. Hence there is a need to modify the rate equations to accurately model finite systems. Effective mesoscopic rate equations (EMRE) have been formulated to account for the breakdown in the linear noise approximation\cite{vankampen, grima1_jcp, grima2_jcp}. Bacteriorhodopsin (bR), a retinal protein is an ideal system to demonstrate the utility of noise studies, as it can be modeled by a 2-state model LEQ in spite of a complex photocycle. The longest time constants are in the ms-s range and can be controlled by external parameters. Finite or small system sizes are essentially defined by the film thickness and the near-field scanning optical microscope (NSOM) probe geometry controlling the excitation volume.

NSOM based sub-diffraction limit imaging technique utilizes narrow optical fiber tips to image single molecules and nano-systems\cite{dunn_cr}. It was shown in our laboratory, that it was possible to monitor the intermediate states of bR, using transmission NSOM\cite{arun_ao}. The light driven proton pump mechanism in bR is initiated by the isomerization of retinal protein from all-trans to 13-cis configuration followed by formation of series of intermediate states, referred to as J, K, L, M, N and O states\cite{stoeck_bb,lanyi_arp}. M state (absorption maxima at wavelength, $\lambda$ $\sim$ 412 nm) representing the deprotonated state has the highest population among all other intermediate states and exhibits strongest spectral shift following photo-excitation at $\lambda$ $\sim$ 570 nm which finally decays thermally or upon blue light illumination. The distinct spectroscopic signature of the intermediate states in the photocycle of bR provides the platform to observe fluctuations associated with the different molecular states. 

To our knowledge, the intensity fluctuations time series in the transmittance of light through a molecular assembly has not been analyzed and related to internal molecular changes. Although single molecule spectroscopic techniques have been demonstrated for fluorescent systems\cite{barkai_arp}. A probable reason could be from the experimental constraint of non-availability of systems with appropriate inherent signal to noise ratio within the data acquisition rates. The advent of NSOM techniques overcomes some of these constraints, and in the process signals from an assembly consisting of $<$ 600 molecules can be closely studied with a combination of probe and pump beams (SN.1)\cite{supple}. A distinct time series signal overriding on a noise feature of sizable amplitude signal describes the transmission fluctuations Tr(t). The characteristic signature of bR in form of pump induced absorption is observed to accompany changes in Tr(t). Tr(t) upon transformation to the frequency domain provides a consistent picture of the events. We implement an algorithm to process the data acquired over large time scales to arrive at a robust representation of the frequency distribution.

\section{THEORY}
\begin{figure*}
\includegraphics[width=14cm]{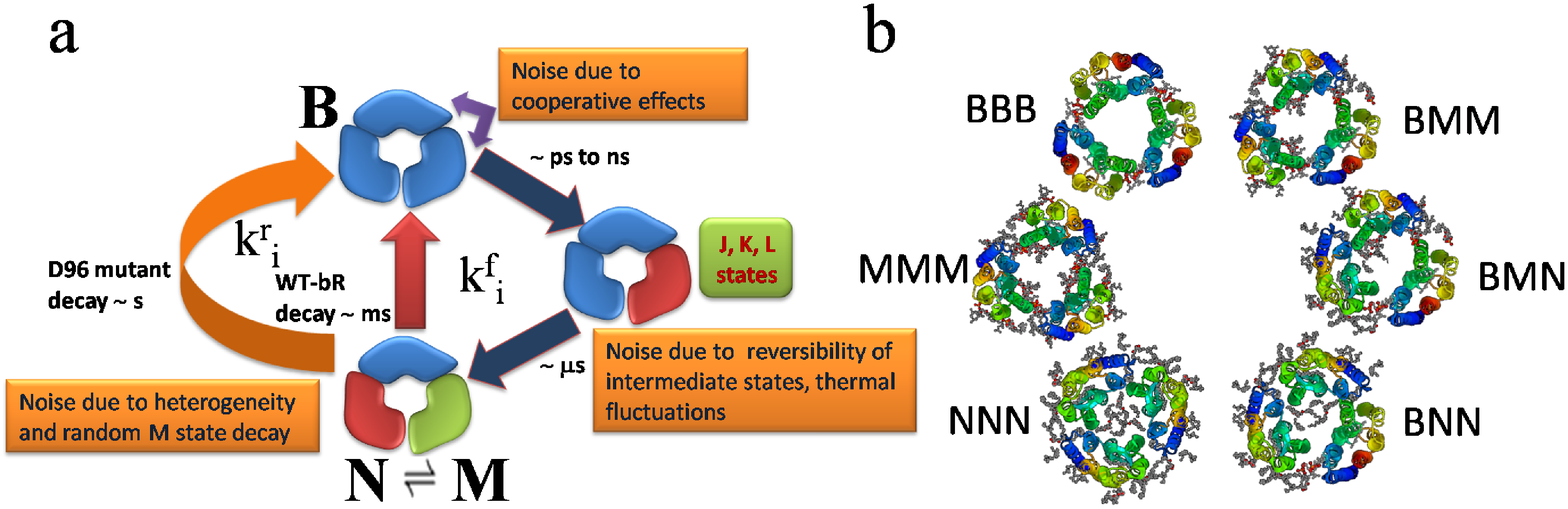}
\caption{Simplified two level energy level scheme of bR showing variability introduced by different sources of noise. (b) Hexagonal arrangement of bR trimers depicting a possible configuration suggesting the statistical source for randomness }\label{two}
\end{figure*}

The kinetics involved in the reaction initiated by the photoexcitaion of the ground state of bR is reasonably well understood. The photocycle kinetics, primarily the M-state lifetimes in bR are known to be affected by pH conditions, additional light-pump at $\lambda_{405}$ corresponding to the M-state excitation, humidity, temperature and presence of metal nanoparticles\cite{sayad_jacs}. These changes introduced by external factors are manifested as peak and linewidth shifts in the noise frequency profile. A quantitative handle on the problem can be initiated using a simplified model where the bR molecule is reduced to a system with two excited states (Fig. \ref{two}a), where B, M and N represents the ground state, excited state, and a long-lived intermediate respectively with a finite reversible rate along with the associated rate constants $k^{f}_{i} (B \rightarrow M)$, $k^{r}_{i} (M \rightarrow B)$ for the $i^{th}$ molecule. $k^{f}_{i}= \alpha(\lambda)I$, where $\alpha(\lambda)$ is the absorption coefficient and I is intensity. $k^{r}_{i}$ on the other hand is the inverse of the M state lifetime ($\tau_{M}$). A set of LEQ can be formulated for the two state model with noise factor as an additional term to justify the nondeterministic outcome: $dN_{B}/dt=-k^{f}_{i} N_{B} + k^{r}_{i} N_{M} + \phi_{B}(t)$ and $dN_{M}/dt=k^{f}_{i} N_{B} - k^{r}_{i} N_{M} + \phi_{M}(t)$ where, $N_{B}$ and $N_{M}$ are concentrations of B and M state of a similar set of $\textit{i}$ molecules and $\phi_{B}$ and $\phi_{M}$ are noise terms. The nature of terms $\phi_{B}$ and $\phi_{M}$ could be white noise with or without correlations. It should be mentioned that the choice of Ornstein Uhlenbeck process with correlations can lead to more realistic solutions. In the presence of $\lambda_{405}$ for exciting the M-intermediate state, these equations get further modified associated with the additional pump-rate. The noise terms in these equations can have a characteristic noise frequency range analogous to the approach by Simpson et al\cite{simpson_jtb}. The PDF can now be attributed to predominantly arise from M state lifetime and can assume a sum of Lorentzian profiles (modified by normal distribution). 

In finite systems the trimers in bR are known not to get simultaneously excited, due to heterogeneity\cite{shibata_nn} (Fig. \ref{two}b). The fluctuations arising from these sources get further modified by the cooperative effects, random excitation and decay of molecules, reversible states in the photocycle and thermally driven transitions. Heterogeneity is averaged out in bulk systems, but in a finite size ensemble they play a significant role in extracting useful molecular information\cite{qua_nm}. The general principles in the method may be applicable to other analogous chromophore containing membrane-protein systems.

\section{MATERIALS AND METHODS}
Wild-type bR (WT-bR) films of different thickness were obtained from aqueous suspension of purple membrane patches ($\sim$ 0.5 mg/ml, pH $\sim$ 9.2)(SF.1,2)\cite{supple, he_am}. These films were characterized by absorption spectroscopy and transient absorption to ascertain bR features (SF.3)\cite{supple}. After the films were prepared on quartz substrates the topography (AFM) (SF.4)\cite{supple} and corresponding transmission (NSOM) map of the sample segment with a probe laser (532 nm) were simultaneously recorded (SF.5) \cite{supple}. Once the bR patch was located, the tip was then positioned onto a point on the patch using software controlled stepper motors and the piezo stage. Typically bR patches have lateral dimensions ranging from 500 nm to 2 $\mu m$ (for multilayer) and heights ranging from 7 nm (monolayer) to several 200 nm (30 layers). The NSOM tip (100 nm diameter) illuminates only a fraction of the region of the 500 nm to 2 $\mu m$ bR patch. The photoexcitaion volume is governed by the aeral coverage of the NSOM tip and the thickness of the patch. A typical Tr(t) trace for WT-bR (Fig. \ref{hist}a (inset)) was obtained using a customized Nanonics MultiView 4000 NSOM model with cantilever tip geometry (SF.6)\cite{supple}. In a typical experiment, a continuous wave laser (probe) was coupled to the NSOM tip ($\sim$ 100 nm) and the transmitted optical signal from a bR patch was measured using a photo-multiplier tube (PMT) through the objective with magnification 50x and N.A = 0.45 (SF.7)\cite{supple}. Additional pump beam (405 nm) could be introduced in the set-up and the measurements with appropriate filters were also carried out (SF.8,9)\cite{supple}. Noise histograms for bare quartz also do not reveal any distinct features when additional pump source is introduced (SF.10)\cite{supple}. The noise histograms for the quartz substrate are primarily a measure of source and detector fluctuations.

Temperature variation was carried out by connecting conducting adhesive-tapes from the scanning stage to a hot plate. Increasing the hot plate temperature resulted in heating the sample stage in the temperature range 296 K to 318 K. The temperature at the sample was calibrated with a digital thermometer. A scan was carried out at room temperature and the transmission data was recorded. Subsequent scans were carried out as the temperature was increased. Humidity variation was carried out by enclosing the scanning stage in a chamber with a steam inlet (SF.11)\cite{supple}. A source of steam was connected and the scan was then carried out. The humidity of the enclosure was independently verified and controlled by the duration of the steam flow into the chamber. Scans were typically carried out in steps of 30 minutes. 

Algorithms used to analyze data were similar to the ones used in photon counting experiments to determine the coherence times of pseudo-thermal sources (SN.2)\cite{ricci_ajp}. Data from PMT was captured using a digital oscilloscope (LeCroy, Wave Runner 6100A) at 250 kHz sampling rate for WT-bR and 500 Hz for D96N mutant. About 100 consecutive data sets of 1 s (for WT-bR) and 10 s (for D96N) were captured as separate windows. For the WT-bR case each window was further split in 15 sub-windows and Fast Fourier Transform (FFT) was applied to each one of these windows. The frequency corresponding to the maximum amplitude in the FFT signal was stored for all the windows. A histogram was constructed of the frequencies and their occurrences leaving out the DC contribution. 

\section{RESULTS}
\begin{figure}
\includegraphics[width=8cm]{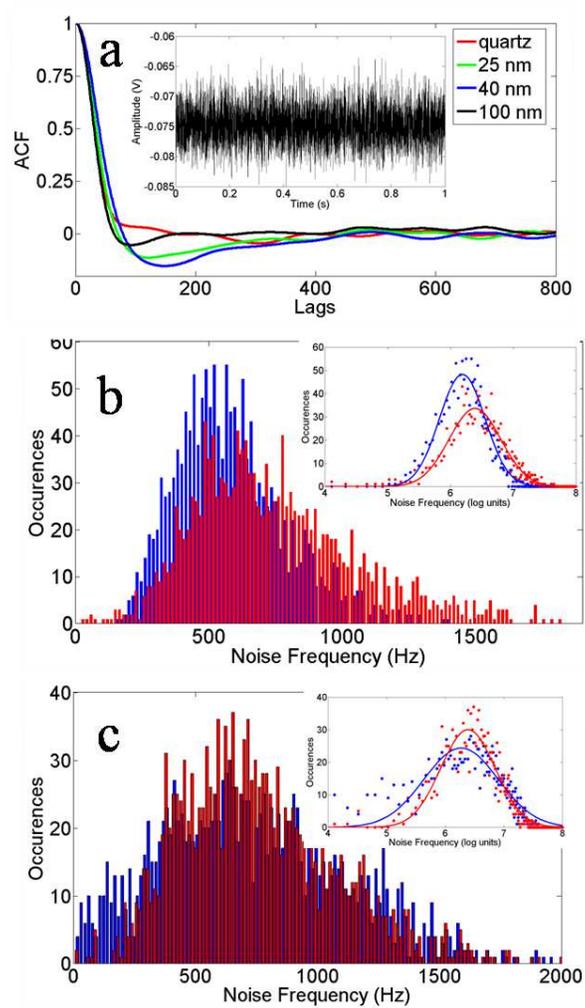}
\caption{(a) Autocorrelation functions for quartz and various bR patches. (inset) Typical time series noise data. Noise power spectra histograms for (b) 40 nm WT-bR patch, (c) 100 nm WT-bR patch, (blue bars, 532 nm illumination, red bars, 532 nm along with pump illumination). (Insets) Corresponding histograms plotted in log frequency scale (blue dots, 532 nm illumination, red dots, 532 nm along with pump illumination,) and Gaussian fits (blue curve, 532 nm illumination, red curve, 532 nm along with pump illumination) }\label{hist}
\end{figure}

These traces are utilized to obtain noise histograms using a rigorous, standardized procedure. Autocorrelation function (ACF) obtained from Tr(t) for quartz and WT-bR samples of different patch heights, clearly indicate the absence of correlations for the bare quartz and a sizable correlation for the bR region (Fig. \ref{hist}a). The frequency distribution arrived from Tr(t) for WT-bR patch was consistently in the form of a lognormal type distribution about a characteristic $\omega_{max}$. Lognormal distribution refers to a skewed distribution where higher frequencies extend the contribution, and whose natural logarithm follows a normal distribution given by the expression, $f(x;\mu,\sigma)= (1/x \sigma \sqrt{(2 \pi)}) (exp(-(ln(x)- \mu )^{2}/(2 \sigma^{2}) )$, where $\mu$ and $\sigma$ are the mean and standard deviation of ln(x) respectively. A common feature arrived upon Tr(t) analysis is the presence of characteristic maximum in the amplitude corresponding to a frequency, $\omega_{max}$ riding a distribution (Fig. \ref{hist}b,c). The magnitude of $\omega_{max} \sim$ 500 - 700 Hz is in the range of the lifetime associated with the M-state which constitutes a large fraction of the entire molecular photocycle span. The interpretation of the dominant frequency associated with the intensity fluctuations to the bR photocycle is consistent with a set of measurements where photocycle rates were intentionally affected by varying certain external factors. The overall profile of the noise spectrum is then expected largely to be controlled by a distribution arising from the heterogeneity in the molecular state in the ensemble and the photocycle kinetics. 

The skewed nature or the non-normal probability density function (PDF) can also point to the statistically correlated nature of these fluctuations (Fig. \ref{hist}b,c). Trimers in bR are known to effects arising from correlations in excitation and de-excitation processes\cite{shibata_nn}. It has been shown that the degree of skewness can be related to the correlation or to the system size\cite{hill_jpc}. It is difficult from our results to quantify the relative contributions from these factors. We speculate that both correlations and finite size effects appear to play an important role  in giving rise to the non-Gaussian PDFs which defines our system.

We do not discount other possible representations such as combination of Lorentzians and Gaussians, but a straight forward lognormal fit yielded reasonable fit parameters. The appropriateness of lognormal function is apparent in form of good Gaussian fits when the distribution is plotted on a log frequency axis (Fig. \ref{hist}b,c inset). Further Kolmogorov-Smirnov tests were done to confirm that the statistics is lognormal (ST.1,2) \cite{supple}. The lognormal behaviour of the distribution is an indicator of statistically correlated events or networks of interacting elements in the present finite size system\cite{koul_jn}. It arises from multiplicative product of many independent random variables.
A shift of $\omega_{max}$ was observed when the additional photo-excitation pump source ($\lambda$ = 405 nm) is introduced in the Tr(t) measurements. It is known that life time of the M state is altered by the thermally driven events in photocycle hastened upon the pumping the intermediate states, thereby biasing the population to the ground state. The blue shift in the $\omega_{max}$ upon $\lambda_{405}$ excitation can then be associated with the reduction in the photocycle span. The parameters in the distribution are also dependent on the number of bR layers (film thickness) (Fig. \ref{hist}b,c). The number of bR layers is a direct way to control the ensemble size. The Tr(t) data exhibits a maximum at 505 Hz for 40 nm thick patch and 630 Hz for 100 nm thick patch which gets shifted to 630 Hz (Fig. \ref{hist}b) and 720 Hz (Fig. \ref{hist}c) upon additional photo-excitation (ST.1)\cite{supple}. Changes in the mean value upon fitting to a lognormal profile, where mean takes up the value of $m=exp(\mu+ \sigma^{2}/2)$, are similar to the changes in $\omega_{max}$ as a function of thickness and pump excitation. It is observed that 25 nm to 150 nm thick patches illuminated by 100 nm tip yields PDFs which reflect internal molecular, M state dependence on external conditions. In case of monolayer which consists of about 200 trimers in the cross section of the incident light beam where quantifiable near-field absorption is observed, the characteristic noise pattern however is missing (SF.9)\cite{supple}. The noise features apparently appear to be dominated by the laser fluctuations and detector characteristics (similar to bare quartz substrates). On the other hand, in the regime of thick films ($>$ 200 nm), heterogeneity and other random molecular events are averaged out and the noise pattern is again featureless\cite{qua_nm}. Hence there seems to be a critical size regime (14 nm - 150 nm) which yields characteristic stochastic features. 

\begin{figure}
\includegraphics[width=8cm]{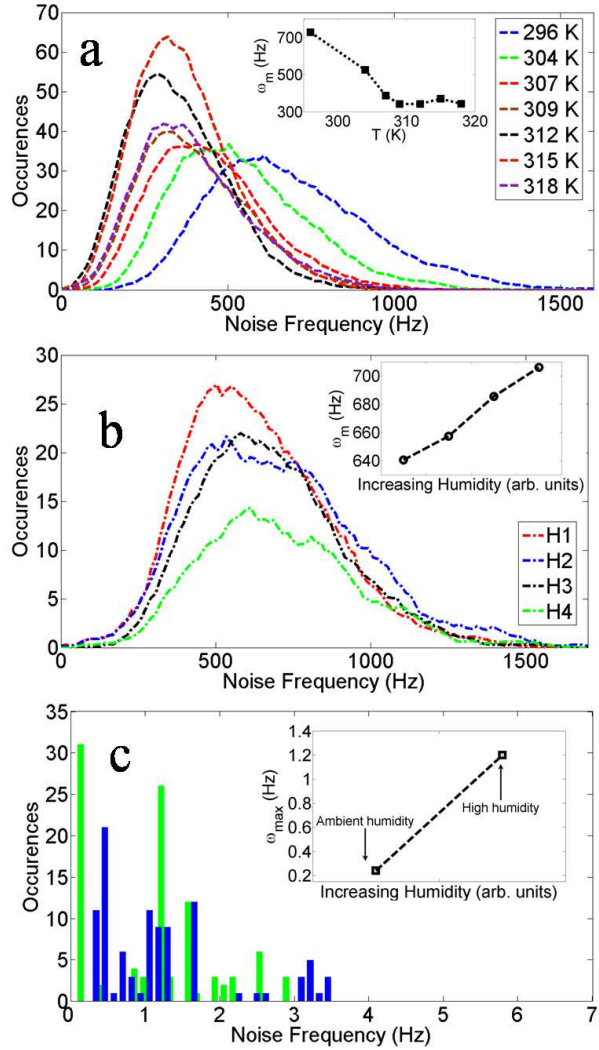}
\caption{(a) Red shift of noise power spectra histograms as a function of temperature for WT-bR. (Inset) Temperature dependence of mean noise frequency for WT-bR. (b) Blue shift of histograms as a function of humidity for WT-bR. (Inset) Humidity dependence of the mean noise frequency (H1$<$H2$<$H3$<$H4). (c) Noise frequency histograms for D96N mutant (green bars, 532 nm illumination, blue bars, 532 nm along with pump illumination). (Inset) Humidity dependence of peak noise frequency for D96N mutant showing variation with humidity.) }\label{comp}
\end{figure}

The T dependent bR photocycle kinetics is modified in dried-film form by the presence of competing factor of moisture concentration related to the proton uptake process . Upon heating the bR films (to 320 K), the increase in moisture depletion with T reduces the photocycle rate. These changes are observed in the form of the red-shift in the $\omega_{m}$ with increasing T (Fig. \ref{comp}a, inset). However, upon increasing the moisture concentration, the expected increase in the photocycle rate in form of the blue shift of $\omega_{m}$ is observed from lognormal distribution fits (Fig. \ref{comp}b, inset). It is interesting to note that the time series data, Tr(t) is also a gauge for ambient conditions, as clearly noted from measurements where the proton concentration gets depleted and enriched upon switching off and on the source for moisture respectively. 
In order to generalize the procedure to other systems we demonstrate the method using genetically modified bR (variant D96N), which has a longer M state lifetime ($>$ 100 ms in film) due to the replacement of aspartic acid (Asp-96) by asparagine (Asn-96)\cite{otto_pnas}. The noise measurements were suitably extended to the low frequency range and the data was collected at 250 KHz and 500 Hz sampling rates. Noise features in D96N: (i) Noise distribution around the maximum of 120 mHz was observed, which shifted to 480 mHz upon the additional $\lambda_{405}$ excitation (Fig. \ref{comp}c). A blue shift was also observed upon increasing the humidity (Fig. \ref{comp}c inset)(SF.12)\cite{supple}. (ii) A frequency distribution was also observed about 750 Hz which however is independent of the illumination. This can correspond to the rise fluctuations which are of the order of ms and dependent only on intensity of illumination and the absorption coefficient (SF.13)\cite{supple}. Low frequency ($<$ 2 Hz) histograms for WT-bR do not exhibit distinct pattern or shifts with pump excitation (SF.14) \cite{supple}.

The Tr(t) measurements on bR films provide an important toolkit as a model system to examine the role and implications of fluctuations. Signature of characteristic noise which appears to be ensemble size dependent in the light-induced proton pump function throws open interesting questions. Molecular processes involved in the functioning of the protein are expected to be dependent on the organization prevailing at much higher length scales and the collective behaviour which emerges depends on these interactions. The interactions can further be probed by varying external parameters and the lognormal distributions analysis of Tr(t) can help quantify the collective response. The general principles in the method should be applicable to other interesting biological systems.

% If in two-column mode, this environment will change to single-column format so that long equations can be displayed. 
%\begin{widetext}
%$$\mbox{put long equation here}$$
%\end{widetext}

% Figures should be put into the text as floats. 
% Use the graphics or graphicx packages (distributed with LaTeX2e).
% Put the label that you will use with \ref{} command in the braces of the \label{} command.
%

% \begin{table}
% \end{tabular}
% \end{table}

\begin{acknowledgments}
We thank Prof. Mudi Sheves for providing the WT-bR suspensions and Prof. David Cahen for help with electrostatic film assembly technique for bR. We acknowledge support from DAE and DST, Govt. of India for partial funding. 
\end{acknowledgments}



\clearpage

\begin{thebibliography}{25}%
\makeatletter
\providecommand \@ifxundefined [1]{%
 \@ifx{#1\undefined}
}%
\providecommand \@ifnum [1]{%
 \ifnum #1\expandafter \@firstoftwo
 \else \expandafter \@secondoftwo
 \fi
}%
\providecommand \@ifx [1]{%
 \ifx #1\expandafter \@firstoftwo
 \else \expandafter \@secondoftwo
 \fi
}%
\providecommand \natexlab [1]{#1}%
\providecommand \enquote  [1]{``#1''}%
\providecommand \bibnamefont  [1]{#1}%
\providecommand \bibfnamefont [1]{#1}%
\providecommand \citenamefont [1]{#1}%
\providecommand \href@noop [0]{\@secondoftwo}%
\providecommand \href [0]{\begingroup \@sanitize@url \@href}%
\providecommand \@href[1]{\@@startlink{#1}\@@href}%
\providecommand \@@href[1]{\endgroup#1\@@endlink}%
\providecommand \@sanitize@url [0]{\catcode `\\12\catcode `\$12\catcode
  `\&12\catcode `\#12\catcode `\^12\catcode `\_12\catcode `\%12\relax}%
\providecommand \@@startlink[1]{}%
\providecommand \@@endlink[0]{}%
\providecommand \url  [0]{\begingroup\@sanitize@url \@url }%
\providecommand \@url [1]{\endgroup\@href {#1}{\urlprefix }}%
\providecommand \urlprefix  [0]{URL }%
\providecommand \Eprint [0]{\href }%
\providecommand \doibase [0]{http://dx.doi.org/}%
\providecommand \selectlanguage [0]{\@gobble}%
\providecommand \bibinfo  [0]{\@secondoftwo}%
\providecommand \bibfield  [0]{\@secondoftwo}%
\providecommand \translation [1]{[#1]}%
\providecommand \BibitemOpen [0]{}%
\providecommand \bibitemStop [0]{}%
\providecommand \bibitemNoStop [0]{.\EOS\space}%
\providecommand \EOS [0]{\spacefactor3000\relax}%
\providecommand \BibitemShut  [1]{\csname bibitem#1\endcsname}%
\let\auto@bib@innerbib\@empty
%</preamble>
\bibitem [{\citenamefont {Chen}\ and\ \citenamefont {Yu}(2007)}]{chen_prl}%
  \BibitemOpen
  \bibfield  {author} {\bibinfo {author} {\bibfnamefont {Z.}~\bibnamefont
  {Chen}}\ and\ \bibinfo {author} {\bibfnamefont {C.~C.}\ \bibnamefont {Yu}},\
  }\bibfield  {title} {\enquote {\bibinfo {title} {Measurement noise maximum as
  a signature of a phase transition},}\ }\href@noop {} {\bibfield  {journal}
  {\bibinfo  {journal} {Phys. Rev. Lett}\ }\textbf {\bibinfo {volume} {98}},\
  \bibinfo {pages} {057204} (\bibinfo {year} {2007})} %
\bibitem [{\citenamefont {Raser}\ and\ \citenamefont
  {O'Shea}(2005)}]{raser_sci}%
  \BibitemOpen
  \bibfield  {author} {\bibinfo {author} {\bibfnamefont {J.~M.}\ \bibnamefont
  {Raser}}\ and\ \bibinfo {author} {\bibfnamefont {E.~K.}\ \bibnamefont
  {O'Shea}},\ }\bibfield  {title} {\enquote {\bibinfo {title} {Noise in gene
  expression: origins, consequences and control},}\ }\href@noop {} {\bibfield
  {journal} {\bibinfo  {journal} {Science}\ }\textbf {\bibinfo {volume}
  {309}},\ \bibinfo {pages} {2010} (\bibinfo {year} {2005})}%
\bibitem [{\citenamefont {Simpson}\ \emph {et~al.}(2009)\citenamefont
  {Simpson}, \citenamefont {Cox}, \citenamefont {Allen}, \citenamefont
  {McCollum}, \citenamefont {Dar}, \citenamefont {Karig},\ and\ \citenamefont
  {Cooke}}]{simpson_nano}%
  \BibitemOpen
  \bibfield  {author} {\bibinfo {author} {\bibfnamefont {M.~L.}\ \bibnamefont
  {Simpson}}, \bibinfo {author} {\bibfnamefont {C.~D.}\ \bibnamefont {Cox}},
  \bibinfo {author} {\bibfnamefont {M.~S.}\ \bibnamefont {Allen}}, \bibinfo
  {author} {\bibfnamefont {J.~M.}\ \bibnamefont {McCollum}}, \bibinfo {author}
  {\bibfnamefont {R.~D.}\ \bibnamefont {Dar}}, \bibinfo {author} {\bibfnamefont
  {D.~K.}\ \bibnamefont {Karig}}, \ and\ \bibinfo {author} {\bibfnamefont
  {J.~F.}\ \bibnamefont {Cooke}},\ }\bibfield  {title} {\enquote {\bibinfo
  {title} {Noise in biological circuits},}\ }\href@noop {} {\bibfield
  {journal} {\bibinfo  {journal} {Nanomed. and Nanobio.}\ }\textbf {\bibinfo
  {volume} {1}},\ \bibinfo {pages} {214} (\bibinfo {year} {2009})}%
\bibitem [{\citenamefont {Cox}\ \emph {et~al.}(2008)\citenamefont {Cox},
  \citenamefont {McCollum}, \citenamefont {Allen}, \citenamefont {Dar},\ and\
  \citenamefont {Simpson}}]{cox_pnas}%
  \BibitemOpen
  \bibfield  {author} {\bibinfo {author} {\bibfnamefont {C.~D.}\ \bibnamefont
  {Cox}}, \bibinfo {author} {\bibfnamefont {J.~M.}\ \bibnamefont {McCollum}},
  \bibinfo {author} {\bibfnamefont {M.~S.}\ \bibnamefont {Allen}}, \bibinfo
  {author} {\bibfnamefont {R.~D.}\ \bibnamefont {Dar}}, \ and\ \bibinfo
  {author} {\bibfnamefont {M.~L.}\ \bibnamefont {Simpson}},\ }\bibfield
  {title} {\enquote {\bibinfo {title} {Using noise to probe and characterize
  gene circuits},}\ }\href@noop {} {\bibfield  {journal} {\bibinfo  {journal}
  {PNAS}\ }\textbf {\bibinfo {volume} {105}},\ \bibinfo {pages} {10809}
  (\bibinfo {year} {2008})} %
\bibitem [{\citenamefont {Rao}, \citenamefont {Wolf},\ and\ \citenamefont
  {Arkin}(2002)}]{rao_nat}%
  \BibitemOpen
  \bibfield  {author} {\bibinfo {author} {\bibfnamefont {C.~V.}\ \bibnamefont
  {Rao}}, \bibinfo {author} {\bibfnamefont {D.~M.}\ \bibnamefont {Wolf}}, \
  and\ \bibinfo {author} {\bibfnamefont {A.~P.}\ \bibnamefont {Arkin}},\
  }\bibfield  {title} {\enquote {\bibinfo {title} {Control, exploitation $\&$
  tolerance of intracellular noise},}\ }\href@noop {} {\bibfield  {journal}
  {\bibinfo  {journal} {Nature}\ }\textbf {\bibinfo {volume} {420}},\ \bibinfo
  {pages} {231} (\bibinfo {year} {2002})} %
\bibitem [{\citenamefont {Grima}(2010{\natexlab{a}})}]{grima_jcp}%
  \BibitemOpen
  \bibfield  {author} {\bibinfo {author} {\bibfnamefont {R.}~\bibnamefont
  {Grima}},\ }\bibfield  {title} {\enquote {\bibinfo {title} {Intrinsic
  biochemical noise in crowded intracellular conditions},}\ }\href@noop {}
  {\bibfield  {journal} {\bibinfo  {journal} {J. Chem. Phys.}\ }\textbf
  {\bibinfo {volume} {132}},\ \bibinfo {pages} {185102} (\bibinfo {year}
  {2010}{\natexlab{a}})} %
\bibitem [{\citenamefont {Montroll}\ and\ \citenamefont
  {Shlesinger}(1982)}]{montroll_pnas}%
  \BibitemOpen
  \bibfield  {author} {\bibinfo {author} {\bibfnamefont {E.~W.}\ \bibnamefont
  {Montroll}}\ and\ \bibinfo {author} {\bibfnamefont {M.~F.}\ \bibnamefont
  {Shlesinger}},\ }\bibfield  {title} {\enquote {\bibinfo {title} {On 1/f noise
  and other distributions with long tails},}\ }\href@noop {} {\bibfield
  {journal} {\bibinfo  {journal} {PNAS}\ }\textbf {\bibinfo {volume} {79}},\
  \bibinfo {pages} {3380} (\bibinfo {year} {1982})} %
\bibitem [{\citenamefont {Hill}, \citenamefont {Dissado},\ and\ \citenamefont
  {Jackson}(1981)}]{hill_jpc}%
  \BibitemOpen
  \bibfield  {author} {\bibinfo {author} {\bibfnamefont {R.~M.}\ \bibnamefont
  {Hill}}, \bibinfo {author} {\bibfnamefont {L.~A.}\ \bibnamefont {Dissado}}, \
  and\ \bibinfo {author} {\bibfnamefont {R.}~\bibnamefont {Jackson}},\
  }\bibfield  {title} {\enquote {\bibinfo {title} {The examination of
  correlated noise},}\ }\href@noop {} {\bibfield  {journal} {\bibinfo
  {journal} {J. Phys. C: Solid State Phys.}\ }\textbf {\bibinfo {volume}
  {14}},\ \bibinfo {pages} {3915} (\bibinfo {year} {1981})}
\bibitem [{\citenamefont {Simpson}, \citenamefont {Cox},\ and\ \citenamefont
  {Sayler}(2004)}]{simpson_jtb}%
  \BibitemOpen
  \bibfield  {author} {\bibinfo {author} {\bibfnamefont {M.~L.}\ \bibnamefont
  {Simpson}}, \bibinfo {author} {\bibfnamefont {C.~D.}\ \bibnamefont {Cox}}, \
  and\ \bibinfo {author} {\bibfnamefont {G.~S.}\ \bibnamefont {Sayler}},\
  }\bibfield  {title} {\enquote {\bibinfo {title} {Frequency domain chemical
  langevin analysis of stochasticity in gene transcriptional regulation},}\
  }\href@noop {} {\bibfield  {journal} {\bibinfo  {journal} {J. of Theo. Bio.}\
  }\textbf {\bibinfo {volume} {229}},\ \bibinfo {pages} {383} (\bibinfo {year}
  {2004})} %
\bibitem [{\citenamefont {Kampen}(2007)}]{vankampen}%
  \BibitemOpen
  \bibfield  {author} {\bibinfo {author} {\bibfnamefont {N.~G.~V.}\
  \bibnamefont {Kampen}},\ }\href@noop {} {\emph {\bibinfo {title} {Stochastic
  Processes in Physics and Chemistry}}}\ (\bibinfo  {publisher} {Elsevier,
  Amsterdam},\ \bibinfo {year} {2007}) %
\bibitem [{\citenamefont {Grima}(2010{\natexlab{b}})}]{grima1_jcp}%
  \BibitemOpen
  \bibfield  {author} {\bibinfo {author} {\bibfnamefont {R.}~\bibnamefont
  {Grima}},\ }\bibfield  {title} {\enquote {\bibinfo {title} {An effective rate
  equation approach to reaction kinetics in small volumes: Theory and
  application to biochemical reactions in nonequilibrium steady-state
  conditions},}\ }\href@noop {} {\bibfield  {journal} {\bibinfo  {journal} {J.
  Chem. Phys.}\ }\textbf {\bibinfo {volume} {133}},\ \bibinfo {pages} {035101}
  (\bibinfo {year} {2010}{\natexlab{b}})} %
\bibitem [{\citenamefont {Thomas}, \citenamefont {Straube},\ and\ \citenamefont
  {Grima}(2010)}]{grima2_jcp}%
  \BibitemOpen
  \bibfield  {author} {\bibinfo {author} {\bibfnamefont {P.}~\bibnamefont
  {Thomas}}, \bibinfo {author} {\bibfnamefont {A.~V.}\ \bibnamefont {Straube}},
  \ and\ \bibinfo {author} {\bibfnamefont {R.}~\bibnamefont {Grima}},\
  }\bibfield  {title} {\enquote {\bibinfo {title} {Stochastic theory of
  large-scale enzyme-reaction networks: Finite copy number corrections to rate
  equation models},}\ }\href@noop {} {\bibfield  {journal} {\bibinfo  {journal}
  {J. Chem. Phys.}\ }\textbf {\bibinfo {volume} {133}},\ \bibinfo {pages}
  {195101} (\bibinfo {year} {2010})} %
\bibitem [{\citenamefont {Dunn}(1999)}]{dunn_cr}%
  \BibitemOpen
  \bibfield  {author} {\bibinfo {author} {\bibfnamefont {R.~C.}\ \bibnamefont
  {Dunn}},\ }\bibfield  {title} {\enquote {\bibinfo {title} {Near-field
  scanning optical microscopy},}\ }\href@noop {} {\bibfield  {journal}
  {\bibinfo  {journal} {Chem.Rev.}\ }\textbf {\bibinfo {volume} {99}},\
  \bibinfo {pages} {2891} (\bibinfo {year} {1999})} %
\bibitem [{\citenamefont {Arun}, \citenamefont {Mukhopadhyay},\ and\
  \citenamefont {Narayan}(2010)}]{arun_ao}%
  \BibitemOpen
  \bibfield  {author} {\bibinfo {author} {\bibfnamefont {N.}~\bibnamefont
  {Arun}}, \bibinfo {author} {\bibfnamefont {S.}~\bibnamefont {Mukhopadhyay}},
  \ and\ \bibinfo {author} {\bibfnamefont {K.~S.}\ \bibnamefont {Narayan}},\
  }\bibfield  {title} {\enquote {\bibinfo {title} {Monitoring intermediate
  states of bacteriorhodopsin monolayers using near field optical
  microscopy},}\ }\href@noop {} {\bibfield  {journal} {\bibinfo  {journal}
  {Appl. Opt.}\ }\textbf {\bibinfo {volume} {49}},\ \bibinfo {pages} {1131}
  (\bibinfo {year} {2010})} %
\bibitem [{\citenamefont {Stoeckenius}, \citenamefont {Lozier},\ and\
  \citenamefont {Bogomolni}(1979)}]{stoeck_bb}%
  \BibitemOpen
  \bibfield  {author} {\bibinfo {author} {\bibfnamefont {W.}~\bibnamefont
  {Stoeckenius}}, \bibinfo {author} {\bibfnamefont {R.~H.}\ \bibnamefont
  {Lozier}}, \ and\ \bibinfo {author} {\bibfnamefont {R.~A.}\ \bibnamefont
  {Bogomolni}},\ }\bibfield  {title} {\enquote {\bibinfo {title}
  {Bacteriorhodopsin and the purple membrane of halobacteria},}\ }\href@noop {}
  {\bibfield  {journal} {\bibinfo  {journal} {Biochem.Biophys. Acta}\ }\textbf
  {\bibinfo {volume} {505}},\ \bibinfo {pages} {215} (\bibinfo {year}
  {1979})} %
\bibitem [{\citenamefont {Lanyi}(2004)}]{lanyi_arp}%
  \BibitemOpen
  \bibfield  {author} {\bibinfo {author} {\bibfnamefont {J.~K.}\ \bibnamefont
  {Lanyi}},\ }\bibfield  {title} {\enquote {\bibinfo {title}
  {Bacteriorhodopsin},}\ }\href@noop {} {\bibfield  {journal} {\bibinfo
  {journal} {Annu. Rev. Physiol.}\ }\textbf {\bibinfo {volume} {66}},\ \bibinfo
  {pages} {665} (\bibinfo {year} {2004})} %
\bibitem [{\citenamefont {Barkai}, \citenamefont {Jung},\ and\ \citenamefont
  {Silbey}(2004)}]{barkai_arp}%
  \BibitemOpen
  \bibfield  {author} {\bibinfo {author} {\bibfnamefont {E.}~\bibnamefont
  {Barkai}}, \bibinfo {author} {\bibfnamefont {Y.~J.}\ \bibnamefont {Jung}}, \
  and\ \bibinfo {author} {\bibfnamefont {R.}~\bibnamefont {Silbey}},\
  }\bibfield  {title} {\enquote {\bibinfo {title} {Theory of single molecule
  spectroscopy: beyond the ensemble average},}\ }\href@noop {} {\bibfield
  {journal} {\bibinfo  {journal} {Annu. Rev. Phys. Chem.}\ }\textbf {\bibinfo
  {volume} {55}},\ \bibinfo {pages} {457} (\bibinfo {year} {2004})}
\bibitem [{sup()}]{supple}%
  \BibitemOpen
  \href@noop {} {}\bibinfo {note} {See supplementary information for
  experimental setup, sample preparation technique, data analysis and
  additional details. [URL will be inserted by AIP]} %
\bibitem [{\citenamefont {Biesso}\ \emph {et~al.}(2009)\citenamefont {Biesso},
  \citenamefont {Qian}, \citenamefont {Huang},\ and\ \citenamefont
  {El-Sayed}}]{sayad_jacs}%
  \BibitemOpen
  \bibfield  {author} {\bibinfo {author} {\bibfnamefont {A.}~\bibnamefont
  {Biesso}}, \bibinfo {author} {\bibfnamefont {W.}~\bibnamefont {Qian}},
  \bibinfo {author} {\bibfnamefont {X.}~\bibnamefont {Huang}}, \ and\ \bibinfo
  {author} {\bibfnamefont {M.~A.}\ \bibnamefont {El-Sayed}},\ }\bibfield
  {title} {\enquote {\bibinfo {title} {Gold nanoparticles surface plasmon field
  effects on the proton pump process of the bacteriorhodopsin
  photosynthesis},}\ }\href@noop {} {\bibfield  {journal} {\bibinfo  {journal}
  {J. Am. Chem. Soc.}\ }\textbf {\bibinfo {volume} {131}},\ \bibinfo {pages}
  {2442} (\bibinfo {year} {2009})} %
\bibitem [{\citenamefont {Shibata}\ \emph {et~al.}(2010)\citenamefont
  {Shibata}, \citenamefont {Yamashita}, \citenamefont {Uchihashi},
  \citenamefont {H.},\ and\ \citenamefont {Ando}}]{shibata_nn}%
  \BibitemOpen
  \bibfield  {author} {\bibinfo {author} {\bibfnamefont {M.}~\bibnamefont
  {Shibata}}, \bibinfo {author} {\bibfnamefont {H.}~\bibnamefont {Yamashita}},
  \bibinfo {author} {\bibfnamefont {T.}~\bibnamefont {Uchihashi}}, \bibinfo
  {author} {\bibfnamefont {H.~K.}\ \bibnamefont {H.}}, \ and\ \bibinfo {author}
  {\bibfnamefont {T.}~\bibnamefont {Ando}},\ }\bibfield  {title} {\enquote
  {\bibinfo {title} {High-speed atomic force microscopy shows dynamic molecular
  processes in photoactivated bacteriorhodopsin},}\ }\href@noop {} {\bibfield
  {journal} {\bibinfo  {journal} {Nat. Nano.}\ }\textbf {\bibinfo {volume}
  {5}},\ \bibinfo {pages} {208} (\bibinfo {year} {2010})} %
\bibitem [{\citenamefont {Quaranta}\ and\ \citenamefont
  {Garbett}(2010)}]{qua_nm}%
  \BibitemOpen
  \bibfield  {author} {\bibinfo {author} {\bibfnamefont {V.}~\bibnamefont
  {Quaranta}}\ and\ \bibinfo {author} {\bibfnamefont {S.~P.}\ \bibnamefont
  {Garbett}},\ }\bibfield  {title} {\enquote {\bibinfo {title} {Not all noise
  is waste},}\ }\href@noop {} {\bibfield  {journal} {\bibinfo  {journal} {Nat.
  Methods}\ }\textbf {\bibinfo {volume} {7}},\ \bibinfo {pages} {269} (\bibinfo
  {year} {2010})} %
\bibitem [{\citenamefont {He}\ \emph {et~al.}(2005)\citenamefont {He},
  \citenamefont {Friedman}, \citenamefont {Cahen},\ and\ \citenamefont
  {Sheves}}]{he_am}%
  \BibitemOpen
  \bibfield  {author} {\bibinfo {author} {\bibfnamefont {T.}~\bibnamefont
  {He}}, \bibinfo {author} {\bibfnamefont {N.}~\bibnamefont {Friedman}},
  \bibinfo {author} {\bibfnamefont {D.}~\bibnamefont {Cahen}}, \ and\ \bibinfo
  {author} {\bibfnamefont {M.}~\bibnamefont {Sheves}},\ }\bibfield  {title}
  {\enquote {\bibinfo {title} {Bacteriorhodopsin monolayers for
  optoelectronics: orientation $\&$ photoelectric response on solid
  supports},}\ }\href@noop {} {\bibfield  {journal} {\bibinfo  {journal} {Adv.
  Mater.}\ }\textbf {\bibinfo {volume} {17}},\ \bibinfo {pages} {1023}
  (\bibinfo {year} {2005})} %
\bibitem [{\citenamefont {Ricci}\ \emph {et~al.}(2007)\citenamefont {Ricci},
  \citenamefont {Mazzaferri}, \citenamefont {Bragas},\ and\ \citenamefont
  {Martinez}}]{ricci_ajp}%
  \BibitemOpen
  \bibfield  {author} {\bibinfo {author} {\bibfnamefont {M.~L.~M.}\
  \bibnamefont {Ricci}}, \bibinfo {author} {\bibfnamefont {J.}~\bibnamefont
  {Mazzaferri}}, \bibinfo {author} {\bibfnamefont {A.~V.}\ \bibnamefont
  {Bragas}}, \ and\ \bibinfo {author} {\bibfnamefont {O.~E.}\ \bibnamefont
  {Martinez}},\ }\bibfield  {title} {\enquote {\bibinfo {title} {Photon
  counting statistics using a digital oscilloscope},}\ }\href@noop {}
  {\bibfield  {journal} {\bibinfo  {journal} {Am. J. Phy.}\ }\textbf {\bibinfo
  {volume} {75}},\ \bibinfo {pages} {707} (\bibinfo {year} {2007})}%
\bibitem [{\citenamefont {Koulakov}, \citenamefont {Hromadka},\ and\
  \citenamefont {Zador}(2009)}]{koul_jn}%
  \BibitemOpen
  \bibfield  {author} {\bibinfo {author} {\bibfnamefont {A.~A.}\ \bibnamefont
  {Koulakov}}, \bibinfo {author} {\bibfnamefont {T.}~\bibnamefont {Hromadka}},
  \ and\ \bibinfo {author} {\bibfnamefont {A.~M.}\ \bibnamefont {Zador}},\
  }\bibfield  {title} {\enquote {\bibinfo {title} {Correlated connectivity and
  the distribution of firing rates in the neocortex},}\ }\href@noop {}
  {\bibfield  {journal} {\bibinfo  {journal} {J. Neuroscience}\ }\textbf
  {\bibinfo {volume} {29}},\ \bibinfo {pages} {3685} (\bibinfo {year}
  {2009})} %
\bibitem [{\citenamefont {Otto}\ \emph {et~al.}(1989)\citenamefont {Otto},
  \citenamefont {Marti}, \citenamefont {Holz}, \citenamefont {Mogi},
  \citenamefont {Lindau}, \citenamefont {Khorana},\ and\ \citenamefont
  {Heyn}}]{otto_pnas}%
  \BibitemOpen
  \bibfield  {author} {\bibinfo {author} {\bibfnamefont {H.}~\bibnamefont
  {Otto}}, \bibinfo {author} {\bibfnamefont {T.}~\bibnamefont {Marti}},
  \bibinfo {author} {\bibfnamefont {M.}~\bibnamefont {Holz}}, \bibinfo {author}
  {\bibfnamefont {T.}~\bibnamefont {Mogi}}, \bibinfo {author} {\bibfnamefont
  {M.}~\bibnamefont {Lindau}}, \bibinfo {author} {\bibfnamefont {H.~G.}\
  \bibnamefont {Khorana}}, \ and\ \bibinfo {author} {\bibfnamefont {M.~P.}\
  \bibnamefont {Heyn}},\ }\bibfield  {title} {\enquote {\bibinfo {title}
  {Aspartic acid-96 is the internal proton donor in the reprotonation of the
  schiff base of bacteriorhodopsin},}\ }\href@noop {} {\bibfield  {journal}
  {\bibinfo  {journal} {PNAS}\ }\textbf {\bibinfo {volume} {86}},\ \bibinfo
  {pages} {9228} (\bibinfo {year} {1989})} %
\end{thebibliography}
\end{document}